\documentclass[twocolumn,english,prl,showpacs,preprintnumbers,superscriptaddress, nofootinbib]{revtex4-2}
\usepackage[utf8]{inputenc}
\usepackage{xcolor}
\usepackage{amsmath}
\usepackage{amssymb}
\usepackage{graphicx}
\usepackage{textcomp}
\usepackage{dcolumn}
\usepackage{bm}
\usepackage[right]{eurosym}
\usepackage{float}
\usepackage[english]{babel}
\usepackage{blindtext}
\usepackage{babel}
\usepackage[normalem]{ulem}

\usepackage[version=3]{mhchem}

\usepackage{ulem}
\definecolor{applegreen}{rgb}{0.55, 0.71, 0.0}

\begin{document}

\title{Direct visualization of an elusive molecular reaction: Time-resolved \ce{H2} roaming in acetonitrile}

\author{Debadarshini Mishra*$^\dagger$}
\affiliation{Department of Physics, University of Connecticut, Storrs, Connecticut, 06269, USA}
\email{These authors contributed equally.}
\author{Aaron C. LaForge*$^\dagger$}
\affiliation{Department of Physics, University of Connecticut, Storrs, Connecticut, 06269, USA}

\author{Lauren M. Gorman}
\affiliation{Department of Physics, University of Connecticut, Storrs, Connecticut, 06269, USA}

\author{Sergio D\'{i}az-Tendero}
\affiliation{Departamento de Qu\'{i}mica, M\'{o}dulo 13, Universidad Aut\'{o}noma de Madrid, 28049 Madrid, Spain, EU}

\affiliation{Condensed Matter Physics Center (IFIMAC), Universidad Aut\'{o}noma de Madrid, 28049 Madrid, Spain, EU}

\affiliation{Institute for Advanced Research in Chemical Sciences (IAdChem), Universidad
Aut\'{o}noma de Madrid, 28049 Madrid, Spain}

\author{Fernando Mart\'{i}n}
\affiliation{Departamento de Qu\'{i}mica, M\'{o}dulo 13, Universidad Aut\'{o}noma de Madrid, 28049 Madrid, Spain, EU}
\affiliation{Condensed Matter Physics Center (IFIMAC), Universidad Aut\'{o}noma de Madrid, 28049 Madrid, Spain, EU}
\affiliation{Instituto Madrile\~no de Estudios Avanzados en Nanociencia (IMDEA-Nano), Campus de Cantoblanco, 28049 Madrid, Spain, EU}

\author{Nora Berrah}
\affiliation{Department of Physics, University of Connecticut, Storrs, Connecticut, 06269, USA}

\def\thefootnote{$\dagger$}\footnotetext{To whom correspondence should be addressed. Email: debadarshini.mishra@uconn.edu or aaron.laforge@uconn.edu}

\begin{abstract}

We introduce a novel approach to imaging the complex phenomenon of roaming, an unconventional mechanism that contributes to molecular dissociation. During roaming, fragments remain weakly-bound to each other due to long-range electrostatic interactions. This extended period of interaction time allows for the formation of new molecular compounds. However, the neutral character of the roaming fragment and its indeterminate trajectory have made it difficult to experimentally identify and systematically study. To address this challenge, we utilize intense, femtosecond IR radiation in combination with coincident Coulomb explosion imaging to directly reconstruct the momentum vector of the neutral roaming \ce{H2} from ionized acetonitrile, \ce{CH3CN}. This technique not only provides a kinematically complete picture of the underlying molecular dynamics that contribute to \ce{H3+} formation, but more significantly, yields an unambiguous experimental signature of roaming. We corroborate this with the aid of quantum chemistry calculations, which allow us to fully determine how this unique dissociative process occurs. Our method enables a direct visualization of roaming reactions, making the investigation of this complex process more feasible.

\end{abstract}

\date{\today}

\maketitle

\section*{Introduction}
Understanding the multitude of complex molecular processes proceeding through photoexcitation is of fundamental importance to a wide variety of biological and chemical systems. Dissociation, one of the more common photochemical processes, typically proceeds by the fragmenting molecule moving along the path of minimum energy. However, there exists a more indirect pathway, known as roaming, where the neutral fragment explores the relatively flat regions of the potential energy surface far from the minimum energy path due to long-range, weakly-bound interactions. As the name suggests, the indirect nature of dissociation through roaming makes this channel more complex than the more typical fragmentation channels. That said, roaming has been a subject of great interest in recent years~\cite{Bowman2011,Bowman2017,Suits2020}. From its initial indirect observation in formaldehyde~\cite{Townsend2004}, roaming has been proposed to occur in several small molecules such as acetaldehyde~\cite{Houston2006,Heazlewood2008}, acetone~\cite{Maeda2010}, nitrate~\cite{Grubb2012}, methyl formate~\cite{Nakamura2015,Lombardi2016}, and propane~\cite{Rauta2016}. Recently, roaming of \ce{H2} in monohydric alcohols, initiated by intense femtosecond laser pulses, was shown to lead to \ce{H3+} formation~\cite{Ekanayake2017,Ekanayake2018,Livshits2020}. In general, photoexcitation of organic molecules, using high-powered, ultrafast lasers, has proven to be an efficient means to produce \ce{H3+}~\cite{Furukawa2005,Hoshina2008,Hoshina2010,Kraus2011,Okino2012,Schirmel2013,Kotsina2015}. Additionally, \ce{H3+}, an abundant cation in interstellar media~\cite{Geballe1996,McCall1998}, is of great interest due to its importance in gas-phase chemistry where it plays a critical role in the production of more complex molecules~\cite{Herbst1973}. 

\begin{figure*}
\centering
\includegraphics[scale=0.45]{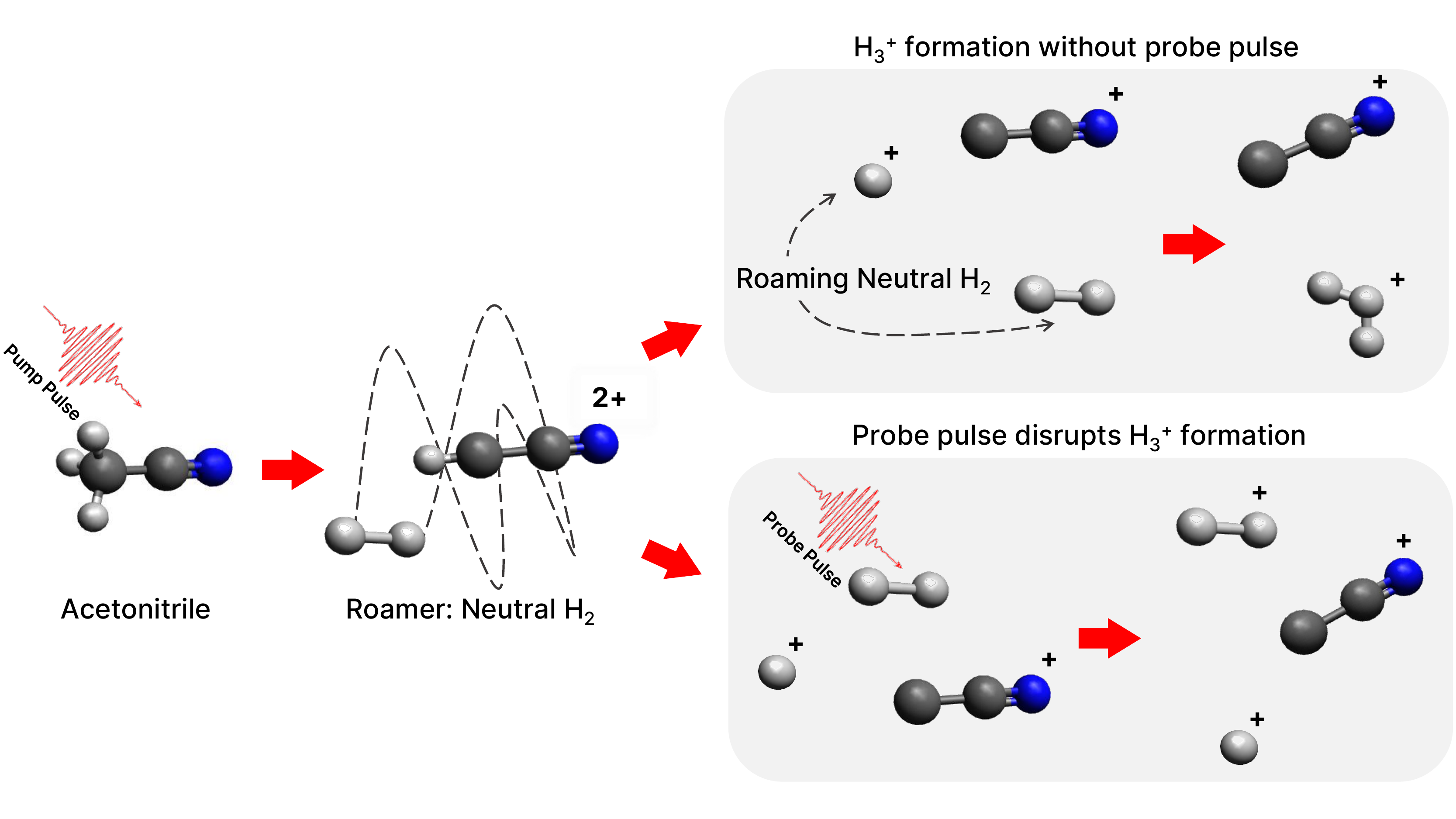}
\caption{Schematic of H$_2$ roaming and H$_3^+$ formation in \ce{CH3CN}. The pump pulse creates $[\ce{CH3CN}]^{2+}$ by multiphoton absorption followed by neutral fragmentation of H$_2$. Some H$_2$ remains weakly-bound to the dication in a roaming state. Afterward, an \ce{H+} is abstracted from the dication forming \ce{H3+}. If the time-delayed probe pulse arrives before abstraction, the process prevents \ce{H3+} formation.}
\label{fig:schematic}
\end{figure*}

Although roaming has been proposed to occur in a multitude of small molecules, a direct visualization of roaming has remained elusive from experimental observations due to its complex nature. The central problem is the random nature of the roaming process itself, where the relative position of the roaming particle with respect to the rest of the system is not deterministic. For instance, pioneering experimental studies ~\cite{Ekanayake2017, Ekanayake2018, Livshits2020} inferred the role of roaming neutral \ce{H2} in the \ce{H3+} formation process by relying heavily on simulations. These studies could not fully exploit the potential of coincident momentum imaging, which is crucial in distinguishing a true experimental signature for roaming from other fragmentation processes. More recently, Endo \textit{et al.}~\cite{Endo2020} demonstrated time-resolved signature of roaming in formaldehyde by investigating a three-ion coincidence channel using Coulomb explosion imaging (CEI). However, this channel carries imprints of various reactions that are triggered via UV excitation of formaldehyde, including molecular and radical dissociation, as well as roaming. As a result, it was necessary to incorporate high-level simulations to identify the signature of roaming from the experimental results.

In this paper, we conduct a systematic, time-resolved investigation of the formation of \ce{H3+} in acetonitrile, \ce{CH3CN}, by directly tracking the dynamics of the roaming neutral \ce{H2}, which is `invisible' to charged-particle detection schemes. We make use of coincident momentum imaging in combination with femtosecond IR-IR pump-probe spectroscopy. This allows us to access the full, time-resolved 3D momentum information of each detected fragment, while also enabling us to reconstruct the momentum vector of the undetected neutral or `invisible' fragment in the reaction, by virtue of momentum conservation. Such kinematically complete information, even for a coincidence channel involving a neutral fragment, is crucial for mapping out a direct signature of roaming reactions, without any interference from other excited state dynamics. Furthermore, the choice of acetonitrile for our current study makes the analysis of \ce{H3+} formation easier, since it only has three hydrogen atoms that are bonded to the same carbon atom, thereby avoiding complications that may arise with molecules that have more than three hydrogen atoms. A schematic of \ce{H2} roaming and \ce{H3+} formation in \ce{CH3CN} is given in Fig.~\ref{fig:schematic}. We produce the parent dicationic state with an intense, femtosecond pump pulse. \ce{H3+} is formed as a result of a roaming \ce{H2} neutral which remains in the flat part of the potential energy surface before abstracting a hydrogen ion from the parent molecule. The formation of \ce{H3+} is probed with a time-delayed fs pulse that disrupts the ensuing dynamics by ionizing the neutral \ce{H2}, thereby preventing the formation of \ce{H3+} and depleting its yield. A more in-depth discussion of our specific experimental setup is given in the Methods section. The momenta of the resulting ionic fragments from the laser-molecule interaction are measured in coincidence using the COLTRIMS technique~\cite{Ullrich1997,Ullrich2003}.

\section*{Results and Discussion}

\begin{figure}
\centering
\includegraphics[width=0.85\linewidth]{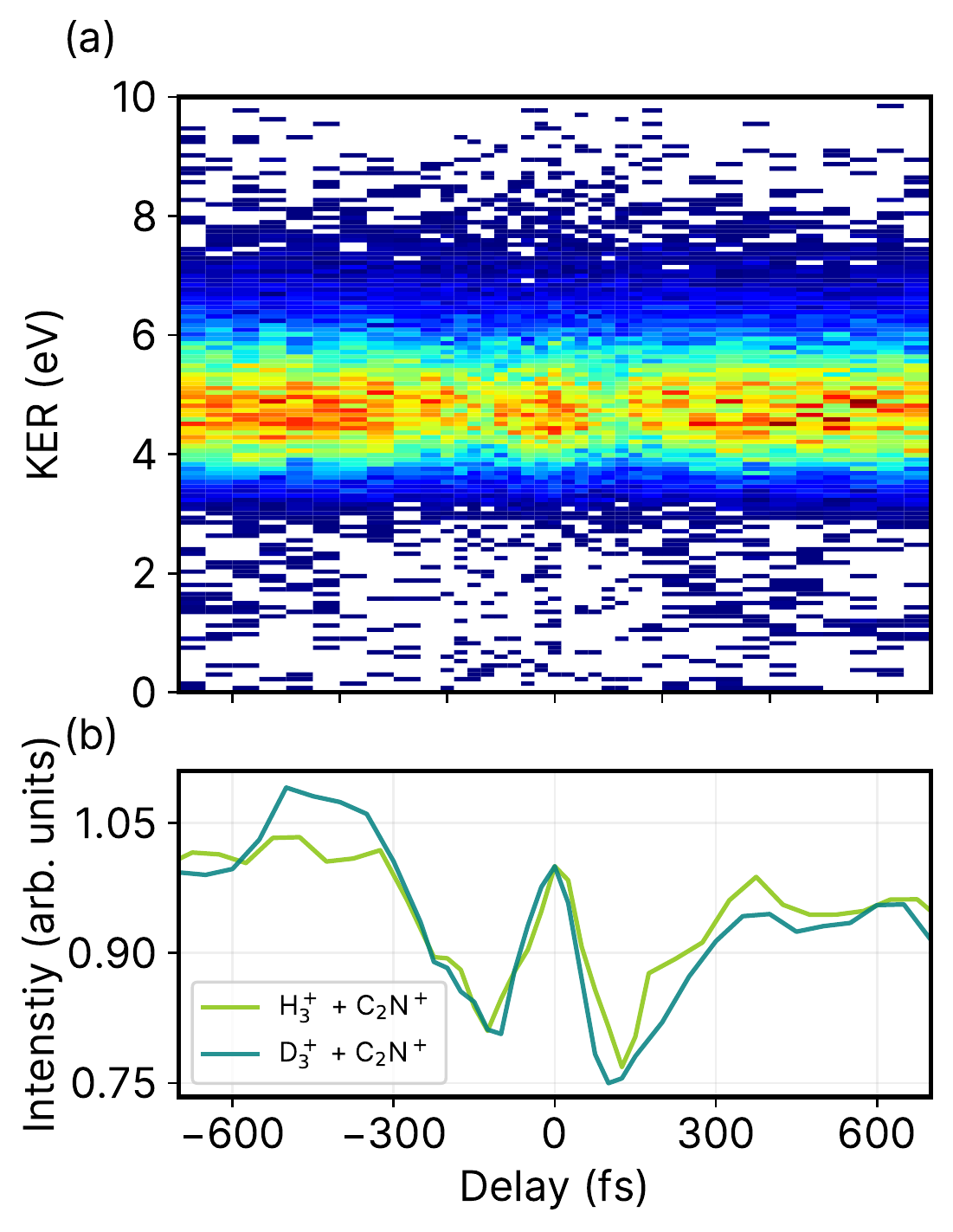}
\caption{(a) Time-dependent kinetic energy release of \ce{H3+} + \ce{C2N+}. (b) Projection of KER signal for \ce{H3+} + \ce{C2N+} \textcolor{applegreen}{(light green)} and \ce{D3+} + \ce{C2N+} \textcolor{teal}{(dark green)} on the pump-probe delay axis. The projected signal intensity in (b) is normalized to 1 at 0 fs pump-probe delay for both \ce{H3+} and \ce{D3+} channels.}
\label{fig:ACNKER}
\end{figure}

To elucidate the role of \ce{H2} roaming in the formation of \ce{H3+} in photoionized acetonitrile, we begin by investigating  its kinematically complete time-resolved dynamics. Fig.~\ref{fig:ACNKER} (a) shows the complete KER map of the channel \ce{H3+} + \ce{C2N+} as a function of pump-probe delay while (b) shows the projection of the KER band between 3 eV and 7 eV. We primarily observe a horizontal band centered around a KER of 5 eV, due to a single pulse exciting the molecule directly into the dicationic state. Typically, these types of KER maps show a dynamic, energy-dependent behavior with respect to pump-probe delay. In that case, the pump pulse excites the system into the ionic state and the time-delayed probe pulse further ionizes the system into a higher charged state, resulting in a mapping of the kinetic energy from Coulomb repulsion as a function of time. In contrast, here we only observe a very weak energy-dependent behavior in the KER map, since the formation of \ce{H3+} is induced directly through the dicationic state. That said, within the horizontal band, there is a strong dependence on the pump-probe delay. From the projection, shown in Fig.~\ref{fig:ACNKER} (b), we see an enhancement in the total yield of this channel around time zero of the pump-probe delay due to the overlap of the two pulses. However, on either side of time zero, we see an immediate decrease in the yield before the signal steadily increases again. As discussed above, this behavior is the result of the disruptive nature of the probe pulse that prevents \ce{H3+} formation by ionizing the neutral roamer \ce{H2} before it can abstract an \ce{H+}. In other words, if the probe pulse arrives early enough, then it `disrupts' the roaming \ce{H2}, resulting in a decreased \ce{H3+} yield. For longer time delays, the probe pulse is less likely to disrupt \ce{H2} roaming; thus, the yield of \ce{H3+} increases until reaching saturation.

We have additionally performed similar measurements using deuterated acetonitrile, which allows us to better resolve different masses when using ion time-of-flight spectroscopy. Overall, we observed no significant differences in the time-resolved measurements for either sample. This, in itself, is a rather interesting result since it reveals that the roaming process relies mostly on electronic properties rather than on nuclear dynamics. This is clearly illustrated in Fig.~\ref{fig:ACNKER} (b), where we show the KER projections of the formation of both \ce{H3+} (light green line) and \ce{D3+} (dark green line). The timescales of depletion and enhancement of the \ce{H3+} and \ce{D3+} projections are almost identical, thus showing that the process is independent of the mass of the roamer. Additionally, we have fitted the projections with a rising exponential of the form: $y=A(1-e^{(t-t_0)/\tau})+y_0$ where $A$ is the normalization constant, $t$ is the pump-probe delay, $t_0$ is the time offset, $\tau$ is the time constant, and $y_0$ is the intensity offset. $A$ and $y_0$ are arbitrary constants since the intensity is normalized to 1 at 0 fs pump-probe delay. However, $t_0$ and $\tau$ have physical significance since they reveal the timescales of the respective depletion and enhancement of the \ce{H3+} and \ce{D3+} formation. The values from the fitting function are given in Table S1 in the supplementary information (SI) while a comparison of the fits with the experimental data is given in Fig. S2 of the SI. The time offset, $t_0$, has an average value of 125 fs, which is larger than the temporal resolution of the experiment (60 fs) and the time constant, $\tau$ has an average value of 105 fs. Overall, the fits reveal that \ce{H2} roaming and \ce{H3+} formation occur on ultrafast timescales of a few hundred femtoseconds.

\begin{figure}
\centering
\includegraphics[width=1\linewidth]{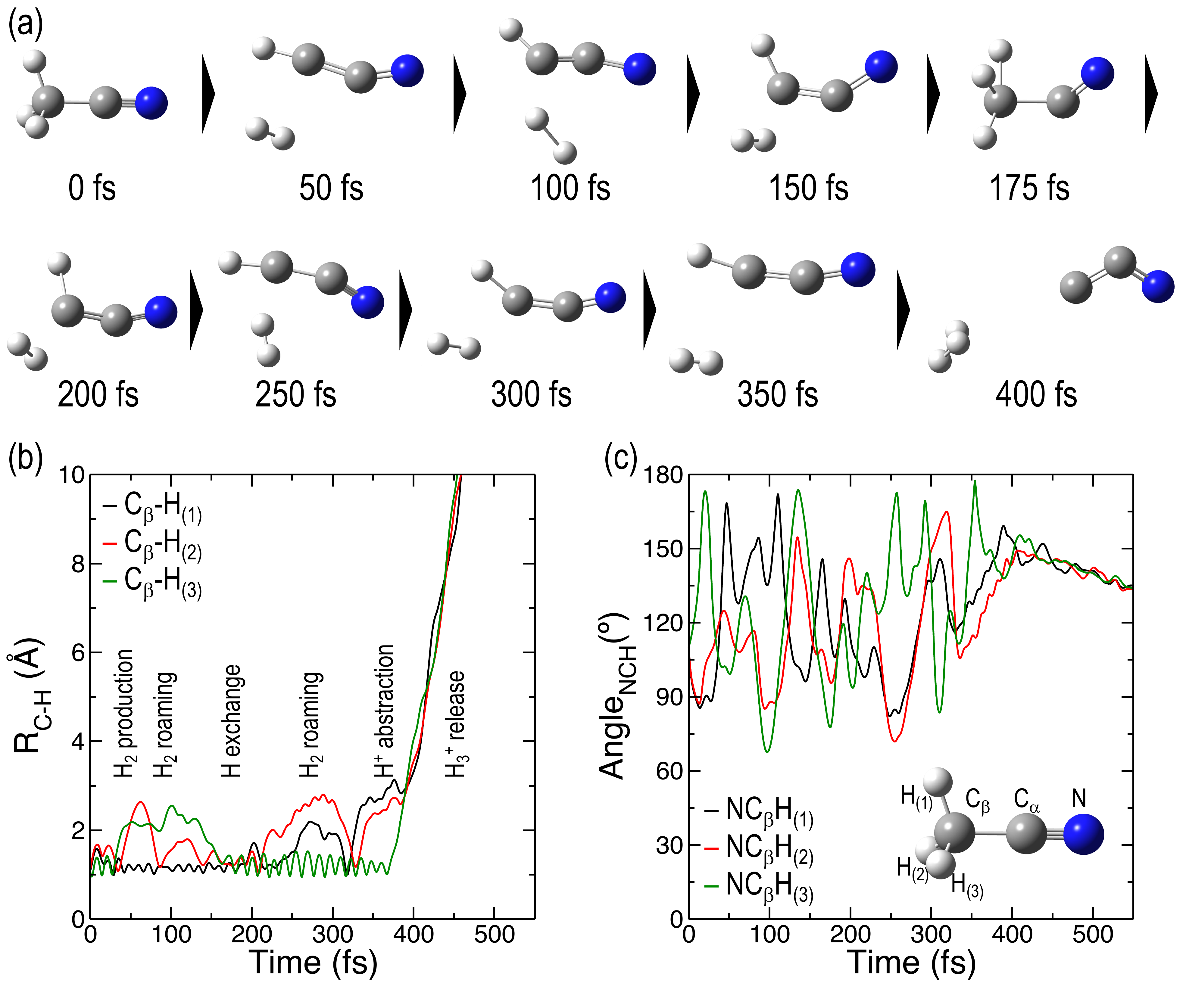}
\caption{(a) Detailed analysis of a trajectory leading to the formation of \ce{H3+}. \ce{H3+} is formed as a result of roaming neutral \ce{H2} followed by a proton abstraction. (b) Time-dependent change of the three C-H bond distances for the same trajectory used in (a), showing \ce{H2} roaming and H abstraction leading to formation of \ce{H3+}. (c) Time-dependent change in the angle between NC$_\beta$ and the three H atoms obtained from the same trajectory used in (a). In this trajectory, the three angles coincide around 400 fs implying ejection of \ce{H3+}. Acetonitrile molecule is shown in the inset identifying the labels for all the carbon and hydrogen atoms.}
\label{fig:H3Trajectory}
\end{figure}

\begin{figure}
\centering
\includegraphics[scale=0.85]{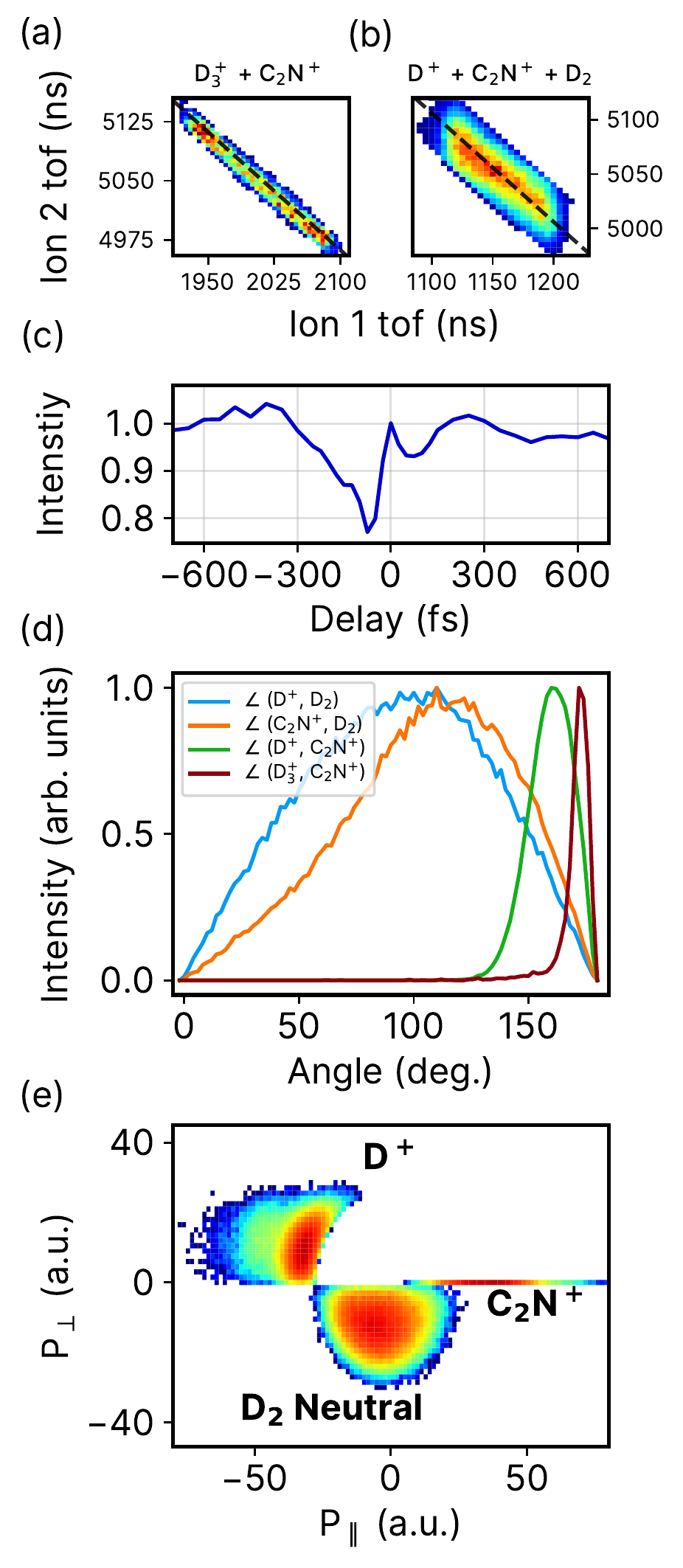}
\caption{Photoion-photoion coincidence lines for (a) \ce{D3+} + \ce{C2N+} and (b) \ce{D+} + \ce{C2N+} + \ce{D2} from acetonitrile, integrated over all pump-probe delays. Black dashed lines of slope -1 are overlaid on the PIPICO channels (a) and (b) to demonstrate the agreement between the expected and experimentally-obtained slope of these channels. (c) Pump-probe delay-dependent kinetic energy projection of the roaming \ce{D2} neutral. (d) Angular distribution between the momentum vectors of \ce{D+} and \ce{D2} \textcolor{blue}{(blue line)}, \ce{C2N+} and \ce{D2} \textcolor{orange}{(orange line)}, \ce{D+} and \ce{C2N+} \textcolor{applegreen}{(green line)}, and \ce{D3+} and \ce{C2N+} \textcolor{purple}{(crimson line)} within the first 200 fs time-delay window. (e) Newton diagram for the channel \ce{D+} + \ce{D2} neutral + \ce{C2N+} integrated over the first 200 fs of pump-probe delay. The momentum vector of \ce{C2N+} lies along the x-axis while those of \ce{D+} and \ce{D2} neutral are plotted in the top and bottom halves, respectively.}
\label{fig:ACNNewton}
\end{figure}

To better understand the mechanism of \ce{H2} roaming and \ce{H3+} formation, we have performed molecular dynamics simulations, described in detail in the Methods section. We have found that the primary fragmentation channels producing cationic hydrogen species and their respective abundances are: 
${\rm H^+/H_2C_2N^+}$ - 79.6\%, ${\rm H_2^+/HC_2N^+}$ - 7.6\%, and ${\rm H_3^+/C_2N^+}$ - 0.8\%. For the current work, we focus the analysis on those trajectories producing ${\rm H_3^+}$. Figure \ref{fig:H3Trajectory} shows the results from one of such trajectories leading to \ce{H3+}. Overall, several processes occur within the first few hundred femtoseconds which impact the  ${\rm H_3^+}$ formation. \ce{H2} is produced in the first tens of fs and then roams around up to $\sim$ 150 fs. Subsequently, H exchange occurs wherein the H atoms of the roaming \ce{H2} molecule are bonded back to the carbon atom, immediately followed by a different pair of H atoms being ejected to form a new roaming \ce{H2} molecule. After another $\sim$ 200 fs, this \ce{H2} molecule abstracts the last remaining H bonded to the C atom, thereby forming \ce{H3+} in coincidence with \ce{C2N+}. The plots in Fig.~\ref{fig:H3Trajectory} (b) and (c) show time-evolution of the bond distances and angles of the three H atoms relative to the molecular skeleton of CCN for a typical trajectory. Results for the other trajectories are given in Figs. S3 and S4, and the corresponding variations of the nuclear kinetic energy with respect to time are given in Fig. S8. In all cases, \ce{H3+} is formed via neutral \ce{H2} roaming, however, due to the small number of trajectories leading to \ce{H3+}, we cannot provide an exact value for the roaming time. That said, all simulated trajectories in the present work predict a formation timescale between 100 fs and 400 fs for \ce{H3+}, which overall agrees very well with the experimental measurements. The calculated nuclear kinetic energies in the final state are also in good agreement with the experimental KER (see Fig. S8).

In order to gain a detailed understanding of the roaming mechanism in acetonitrile, we track the dynamics of the neutral \ce{D2} which appears in coincidence with \ce{D+} and \ce{C2N+} using photoion–photoion coincidence (PIPICO) maps. In general, negative-sloping distributions observed in the PIPICO coincidence maps identify fragments created by multiple ionization. Specifically, the shape and slope of the distribution give information about the dissociation process~\cite{Eland1991}. For instance, the back-to-back emission of two ions, where momentum is conserved, results in a sharp line of slope -1. For deuterated acetonitrile, this is observed in Fig.~\ref{fig:ACNNewton} (a) for \ce{D3+} in coincidence with \ce{C2N+}. Prior to \ce{D3+} formation, the PIPICO channel that allows us to track the behavior of the neutral roamer is the incomplete channel of \ce{D+} in coincidence with \ce{C2N+}, shown in Fig.~\ref{fig:ACNNewton} (b), which is missing the mass from \ce{D2}. Incomplete coincidence channels can have unique shapes and slopes depending on the missing fragment~\cite{Eland1991}. 
When there is deferred charge separation of the different fragments with mass $m$ such that a neutral $m_3$ is ejected first, followed by $m_1m_2^{2+}$ breaking apart into two ions, the resulting coincidence channel has a broad, lozenge shape of slope -1. This is precisely what we observe in Fig.~\ref{fig:ACNNewton} (b) for the channel \ce{D+} + \ce{C2N+} + \ce{D2}. To show the agreement between the calculated slopes and experimental data, black dashed lines of slope -1 are overlaid in the PIPICO channels in Figs.~\ref{fig:ACNNewton} (a) and (b), respectively. To further prove that the missing fragment in this channel is indeed neutral and not an ion that was simply lost in the detection process, we have plotted the kinetic energy of the missing \ce{D2} in Fig. S5 of the SI. If it is an ion, one would observe a time-dependent decrease in its kinetic energy as a result of Coulomb explosion from the remaining charged moiety. On the other hand, a neutral dissociating from a charged fragment would result in very low, time-independent kinetic energy. The latter is observed in the reconstructed kinetic energy of \ce{D2}, thus proving that the missing fragment is indeed neutral. Fig.~\ref{fig:ACNNewton} (c) shows the reconstructed kinetic energy projection of neutral \ce{D2} as a function of pump-probe delay where the time-dependent behavior of this neutral fragment is similar to that of \ce{H3+} + \ce{C2N+}, shown in Fig.~\ref{fig:ACNKER} (b).

After isolating the coincidence channel involving the target neutral roamer \ce{D2}, we then study its angular correlations with respect to the coincident ionic fragments. In Fig.~\ref{fig:ACNNewton} (d), we plot the angle between the momentum vectors of \ce{D2} and \ce{D+} (blue line), and \ce{D2} and \ce{C2N+} (orange line). The momentum vector of \ce{D2}, which is invisible to our detection scheme, is reconstructed by implementing momentum conservation among the three fragments. Overall, we see a very broad distribution implying a highly uncorrelated emission angle between the neutral \ce{D2} and the two ionic fragments, which is a key signature of roaming. Unlike direct dissociation where the fragments follow a fixed minimum energy path, roaming fragments traverse relatively flat regions of the potential energy surface, which results in wide variations in their pathways. Such an uncorrelated behavior is exactly what makes roaming difficult to observe since the position of the roaming neutral is not well-defined as compared to a typical dissociation channel. In that regard, we have a potentially more direct method for visualizing roaming by directly mapping the momentum of the reconstructed neutral. 
For comparison, the ion-ion angular distribution for \ce{D+} and \ce{C2N+} (green line) in this incomplete channel is much sharper and centered at $\sim$ 160$^\circ$, indicating strong Coulomb interaction resulting in a stronger momentum correlation. Such an angular distribution between \ce{D+} and \ce{C2N+} is also observed from the calculated roaming trajectories. For the trajectory shown in Fig.~\ref{fig:H3Trajectory} (c), the angle between \ce{D+} and \ce{C2N+} in the first 200 fs is approximately 150$^\circ$ with a dispersion around this value of the order of 30$^\circ$. Finally, the angular distribution for \ce{D3+} and \ce{C2N+} (crimson line), which forms a complete two-body channel, is even narrower and centered at $\sim$ 175$^\circ$ indicating back-to-back emission, as would be expected from a two-body fragmentation process. 

Beyond examining simple angular correlations, we can also construct the full 3D momentum of the reaction, using Newton diagrams. In general, Newton diagrams are very useful tools for providing structural information about the reaction dynamics of a system~\cite{Boll2022, Bhattacharyya2022, Pathak2020}. One of their unique features is the ability to distinguish different mechanisms of fragmentation. For instance, concerted, instantaneous fragmentation would appear as sharp, localized distributions in the Newton diagram~\cite{Xie2015}, whereas sequential fragmentation would appear as a semicircle~\cite{Neumann2010, Pathak2020} due to the rotation of the intermediate fragment whose lifetime is larger than its rotational period. A typical Newton diagram for a three-body fragmentation channel in acetonitrile, \ce{D+} + \ce{D2+} + \ce{C2N+}, is shown in Fig. S6 of the SI. In this case, the momentum distributions are sharp with strong angular correlations.

Figure~\ref{fig:ACNNewton} (e) shows the Newton diagram for the channel \ce{D+} + \ce{D2} neutral + \ce{C2N+}. Here, the pump-probe delay is constrained to values less than 200 fs. The momentum vector of \ce{C2N+} is fixed along the x-axis, while those of \ce{D+} and the reconstructed \ce{D2} are plotted on the top and bottom halves, respectively. Due to momentum conservation amongst the three fragments, the Newton diagram lies on a plane and shows the relative momentum distributions of two fragments with respect to \ce{C2N+}. Overall, there are several features that are markedly different from the analogous all-ion channel (\ce{D+} + \ce{D2+} + \ce{C2N+}), mentioned above. The distribution of the neutral \ce{D2} in Fig.~\ref{fig:ACNNewton} (e) appears quite different from a typical fragmentation channel since there is no clear angular dependence in addition to a very broad, albeit lower momentum magnitude distribution. This shows that the neutral \ce{D2} departs with a smaller fraction of the total momentum involved in this process, as would be expected from neutral fragmentation. It must be noted that, in sequential processes, rotation of the intermediate fragment can also result in broad angular distributions. Therefore, the Newton diagram in Fig.~\ref{fig:ACNNewton} (e) is integrated over the first 200 fs pump-probe delay window, which is much smaller than the rotational timescale for \ce{C2ND+}. Thus, the broad angular correlation that we observe here cannot be due to the rotation of the intermediate fragment \ce{C2ND+}, thereby removing any ambiguity between rotation and roaming. Such a lack of clear angular correlation in conjunction with a broad distribution of momentum magnitude of the neutral \ce{D2} is a potentially unambiguous signature for roaming in molecules. Contrary to that, the two ions in this incomplete channel, \ce{D+} and \ce{C2N+}, share a stronger momentum correlation mediated via Coulomb interaction as evidenced by a narrower angular spread along with a localized distribution in the magnitude of momentum.

\begin{figure}
\centering
\includegraphics[width=1\linewidth]{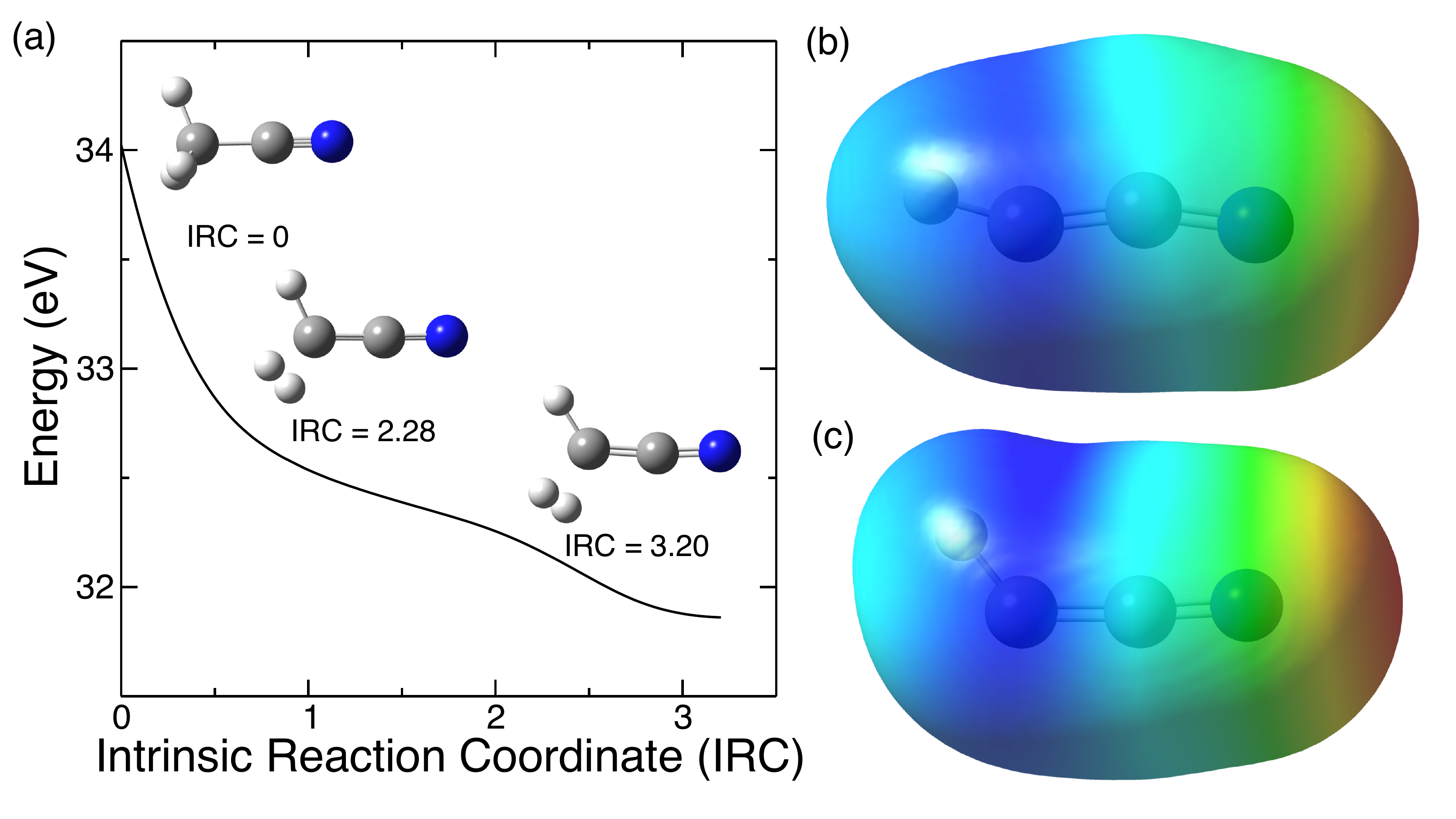}
\caption{(a) Minimum Energy Path (MEP) followed in the potential energy surface of the doubly ionized acetonitrile; relative energy in eV referred to the neutral molecule as a function of the Intrinsic Reaction Coordinate (IRC). (b) and (c) Electrostatic potential (ESP) of $\rm HCCN^{2+}$ mapped on the electronic density with isovalue 0.0004 a.u.. Color code for the extreme values: red = 0.32 a.u. and dark blue = 0.48 a.u.. The geometry used in (b) corresponds to the channel $\rm H_2/HCCN^{2+}$, i.e. after release of neutral $\rm H_2$, and in (c) to the weakly bonded $\rm H_2...HCCN^{2+}$, i.e. last point in the MEP.}
\label{fig:roamingpotential}
\end{figure}

To demonstrate how robust this method is for distinguishing neutral fragmentation channels, in Fig. S1 of the SI, we compare our results to a different type of neutral fragmentation, which was measured in 2-propanol(\ce{CH3CHOHCH3}). In this case, the fragmentation channel, \ce{CH3+} + \ce{H2O} neutral + \ce{C2H3+}, is initially two primary fragments, \ce{CH3+} and \ce{C2H5O+}, the latter of which subsequently dissociates into \ce{C2H3+} and neutral \ce{H2O}. In comparison to neutral roaming, we observe stark differences in the PIPICO channel, angular distributions, and Netwon diagram, which are discussed in detail in the SI.  

Further insight into \ce{H2} roaming has been obtained in a careful exploration of the potential energy surface (PES). In particular, we computed the minimum energy path (MEP) that a doubly charged acetonitrile molecule would follow after vertical ionization, i.e., the maximum gradient on the PES, starting from the optimized geometry of the neutral system (Fig.~\ref{fig:roamingpotential}(a)). We clearly observe the release of neutral $\rm H_2$, followed by the production of a weakly bonded $\rm H_2...HCCN^{2+}$ complex. Furthermore, this shows that the neutral $\rm H_2$ is polarized by the potential exerted by the cationic fragment. We evaluated such potential by computing the electrostatic potential (ESP) of $\rm HCCN^{2+}$ in two different configurations: using the geometry of the optimized structure from the moiety (Fig.~\ref{fig:roamingpotential}(b)) and using the geometry of the last point given in the MEP exploration (Fig.~\ref{fig:roamingpotential}(c)). The ESP is plotted in both cases by projecting the actual value on the electronic density isosurface, which provides a visualization of the potential that polarizes the neutral $\rm H_2$. Overall, Figs.~\ref{fig:roamingpotential} (b) and (c) reveal that the regions with the highest electrostatic potential (blue in color) are nearest to the C--H bond in $\rm HCCN^{2+}$. It is most likely that the roaming \ce{H2} remains weakly-bound in this region. Upon fragmentation of \ce{H+} and \ce{CCN+}, the \ce{H2} is approximately at 90$^\circ$ with respect to the two ionic fragments, which qualitatively agrees with the angular correlations given in Fig.~\ref{fig:ACNNewton}(d).

The fact that acetonitrile has only 3 hydrogen atoms that can contribute to \ce{D3+} formation makes it an ideal candidate to unambiguously track such roaming reactions. However, in order to be exhaustive in terms of all the possible ways that \ce{D3+} can be formed in acetonitrile, we must also consider the possibility of a roaming neutral \ce{D} which can abstract \ce{D2} from the remaining moiety to form \ce{D3+}. 
For this, we have to consider the incomplete coincidence channel: \ce{D2+} + \ce{C2N+} + \ce{D}. We note that the yield for this channel is nearly 4.5 times lower than \ce{D+} + \ce{C2N+} + \ce{D2}, consequently making it a weak contributor to the \ce{D3+} channel. 
This is not surprising, since a free \ce{H} or \ce{D} atom is a radical, which makes it less likely to be formed than \ce{H2} or \ce{D2}.
In fact, this channel is not observed in our theoretical simulations due to the limited number of trajectories used in the present work. To our knowledge, it has also not been observed in previous works. Following the same sequence as before, we first ensure that the missing \ce{D} is indeed a neutral by looking at its kinetic energy as a function of time-delay. Furthermore, in Fig. S7, we show the angular correlation between all pairs of fragments in the channel \ce{D2+} + \ce{D} neutral + \ce{C2N+}. The broad and symmetric angular correlations between the neutral \ce{D} and the other two ionic fragments are indications of roaming \ce{D}. Unlike in Fig.~\ref{fig:ACNNewton}, the correlation between \ce{D} and \ce{C2N+} is quite symmetric. This could be a result of \ce{D} being a lighter roaming fragment compared to \ce{D2}.

\section*{Conclusion}
In conclusion, we have developed a more direct means to image neutral roaming reactions in small molecular systems. Using ultrafast IR pump-IR probe spectroscopy combined with coincident Coulomb explosion imaging, we have time-resolved the formation of \ce{H3+} from neutral \ce{H2} roaming in acetonitrile, which was measured to occur within a few hundred femtoseconds. This technique enables us to directly track the `invisible' neutral roamer, the precursor to \ce{H3+} formation, from incomplete fragmentation channels. Additionally, with the aid of quantum chemistry calculations, we have fully simulated the possible roaming trajectories, allowing us to follow some of the unique intramolecular processes which occur in this type of dissociation. In general, our novel technique gives us a more straightforward means to observe neutral fragments which can allow us to gain a better understanding of the underlying molecular dynamics in roaming reactions.

\section*{Methods}
The experimental setup has been previously described elsewhere~\cite{Kling2019,McDonnell2020,Mishra2022}.
We used a 5 kHz Ti:Sapphire laser producing 35 fs pulses with a central wavelength of 800 nm. 
The laser beam was split into two independent arms which were time-delayed with respect to one another. Each arm had a variable intensity that was optimized to produce the highest contrast on the \ce{H3+/D3+} signal. For this experiment, the power was in the range of $10^{14}$ W/cm$^2$.
The two beams were directed into the interaction region where they were overlapped and back-focused to a spot size of about 10 $\mu$m with a temporal resolution of about 60 fs. 
Both beams were linearly polarized in the direction of the time-of-flight axis of the spectrometer. 
A cold, molecular jet of acetonitrile, from a room-temperature bubbler, was produced by expansion through a 30 $\mu$m nozzle seeded with 1 bar of helium gas and propagated into the spectrometer perpendicular to the two laser pulses. 
Acetonitrile was ionized and resulting charged fragments were detected by a Cold Target Recoil Ion Momentum Spectrometer (COLTRIMS)~\cite{Ullrich1997,Ullrich2003}. 
Using a weak, homogeneous electric field, the ions were directed towards a position-sensitive detector, which is capable of measuring the three-dimensional momentum distributions of the charged particles. 
For CEI, we primarily focus on measuring the fragmented ions, and for the current experimental conditions, we have an ion momentum resolution of 0.1 a.u. 
By applying the coincidence technique, we can isolate specific mass channels of acetonitrile and gain the most relevant information about the fragmentation dynamics.

\emph{Ab initio} molecular dynamics simulations were performed, using the Atom Centered Density Matrix Propagation method (ADMP) \cite{Schlegel2001,Iyengar2001,Schlegel2002} as implemented in the Gaussian16 program \cite{Gaussian16}. The electronic structure was computed, employing the density functional theory (DFT), in particular, the B3LYP functional \cite{Becke1993,Lee1988} in combination with the 6-31++G(d,p) basis set \cite{Clark1983,Krishnan1980}. To ensure the adiabaticity of the dynamics, we established a time step of $\Delta t$ = 0.1 fs and a fictitious electron mass of $\mu$ = 0.1 amu. Mimicking the experimental conditions, we considered a vertical double ionization of the acetonitrile molecule in a Franck–Condon manner, introducing a given amount of internal energy, $E_{exc}$ = 3 eV, which was randomly redistributed over the nuclear degrees of freedom. This value was chosen to ensure that the energy available in the system agrees with the experimentally measured KER. 500 trajectories were propagated up to 1ps. We analyzed the statistical population of the different channels considering separated fragments with distances larger than 2.5 \AA\; between atoms forming a moiety.
Charges in the different atomic positions were computed from a Mulliken population analysis in the last step of each trajectory.
We have successfully used this computational strategy in the past to infer the fragmentation dynamics in ionized molecules and clusters in the gas phase (see e.g. \cite{Maclot2013,Kukk2015,Piekarski2015,Maclot2016,Kling2019,McDonnell2020,Rousseau2020,Barreiro2021,Mishra2022,Ganguly2022}).

\section*{Acknowledgements}
The experimental work was funded by the National Science Foundation under award No. 1700551. The theory was supported by the MICINN (Spanish Ministry of Science and Innovation) projects PID2019-105458RB-I00 and PID2019-110091GB-I00, funded by MCIN/AEI/10.13039/501100011033, the ‘Severo Ochoa’ Programme for Centres of Excellence in R\,$\&$\,D (CEX2020-001039-S) and the ‘Mar\'{i}a de Maeztu’ Programme for Units of Excellence in R\,$\&$\,D (CEX2018-000805-M). We acknowledge the generous allocation of computer time at the Centro de Computaci\'{o}n Cient\'{i}fica at the Universidad Aut\'{o}noma de Madrid (CCC-UAM).

\bibliography{acetonitrile}

\end{document}


\title{Supplemental Information for -- Direct visualization of an elusive molecular reaction: Time-resolved \ce{H2} roaming in acetonitrile}

\author{Debadarshini Mishra*$^\dagger$}
\affiliation{Department of Physics, University of Connecticut, Storrs, Connecticut, 06269, USA}

\author{Aaron C. LaForge*$^\dagger$}
\affiliation{Department of Physics, University of Connecticut, Storrs, Connecticut, 06269, USA}

\author{Lauren M. Gorman}
\affiliation{Department of Physics, University of Connecticut, Storrs, Connecticut, 06269, USA}

\author{Sergio D\'{i}az-Tendero}
\affiliation{Departamento de Qu\'{i}mica, M\'{o}dulo 13, Universidad Aut\'{o}noma de Madrid, 28049 Madrid, Spain, EU}

\affiliation{Condensed Matter Physics Center (IFIMAC), Universidad Aut\'{o}noma de Madrid, 28049 Madrid, Spain, EU}

\affiliation{Institute for Advanced Research in Chemical Sciences (IAdChem), Universidad
Aut\'{o}noma de Madrid, 28049 Madrid, Spain}

\author{Fernando Mart\'{i}n}
\affiliation{Departamento de Qu\'{i}mica, M\'{o}dulo 13, Universidad Aut\'{o}noma de Madrid, 28049 Madrid, Spain, EU}
\affiliation{Condensed Matter Physics Center (IFIMAC), Universidad Aut\'{o}noma de Madrid, 28049 Madrid, Spain, EU}
\affiliation{Instituto Madrile\~no de Estudios Avanzados en Nanociencia (IMDEA-Nano), Campus de Cantoblanco, 28049 Madrid, Spain, EU}

\author{Nora Berrah}
\affiliation{Department of Physics, University of Connecticut, Storrs, Connecticut, 06269, USA}

\maketitle

\section*{Neutral Fragmentation Channel in 2-propanol}

\begin{figure}[H]
\centering
\includegraphics[scale=0.65]{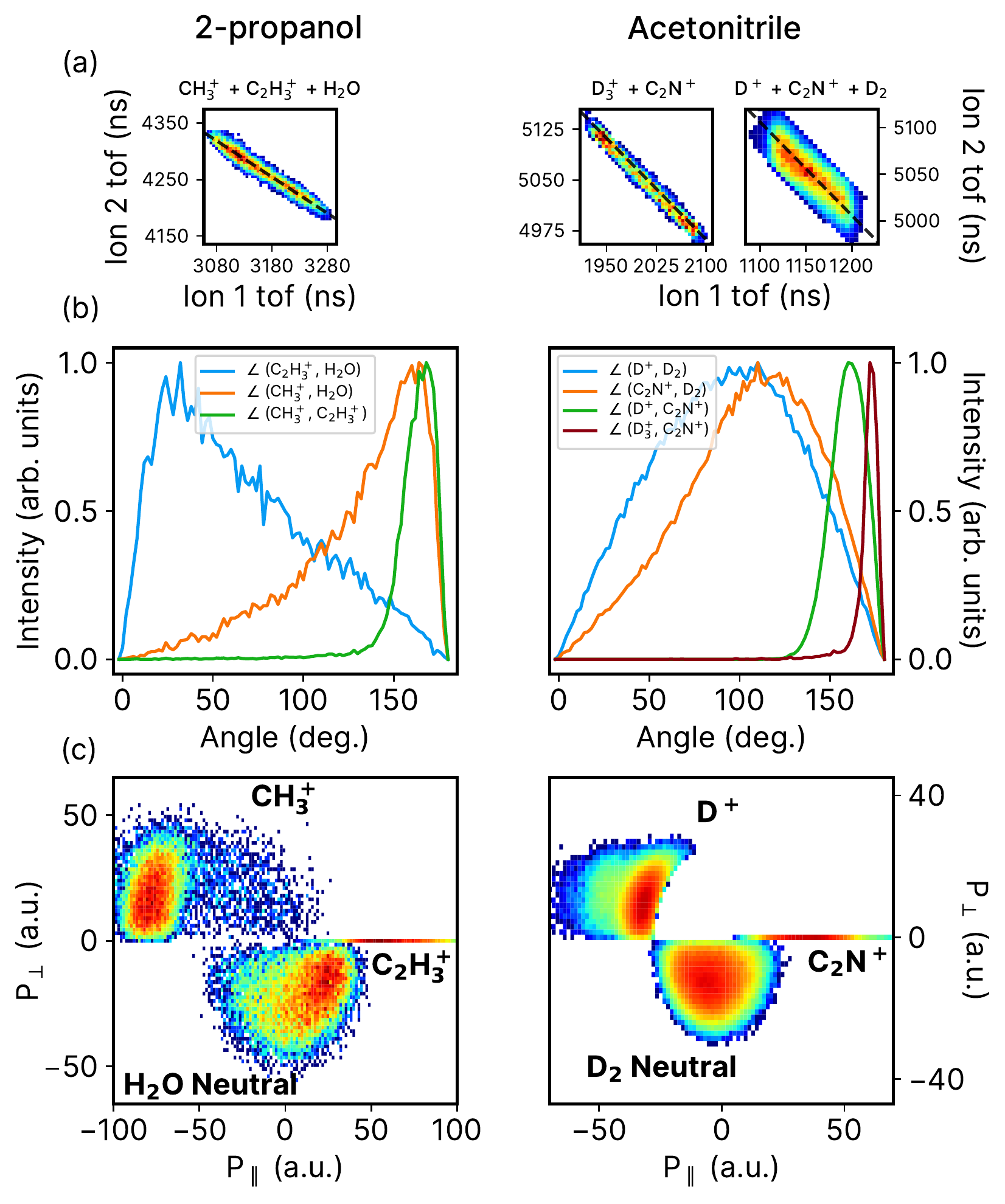}
\caption{(a) Incomplete PIPICO channel for \ce{CH3+} + \ce{H2O} neutral + \ce{C2H3+} from 2-propanol integrated over the first 200 fs. A black dashed line of slope -0.643 is overlaid on the PIPICO channel to show the agreement between the calculated slope and experimental coincidence channel. (b) Angular distributions between the momentum vectors of \ce{C2H3+} and \ce{H2O} \textcolor{blue}{(blue line)}, \ce{CH3+} and \ce{H2O} \textcolor{orange}{(orange line)}, \ce{CH3+} and \ce{C2H3+} \textcolor{applegreen}{(green line)} within the first 200 fs time-delay window. (c) Newton plot for the channel \ce{CH3+} + \ce{C2H3+} + \ce{H2O} neutral integrated over the first 200 fs of pump-probe delay. The momentum vector of \ce{C2H3+} lies along the x-axis while those of \ce{CH3+} and \ce{H2O} neutral are plotted in the top and bottom halves, respectively. Fig. 4 from the main text is shown in the right column for comparison to acetonitrile.}
\label{fig:3BodyIsoPropanol}
\end{figure}

In order to fully demonstrate the differences between a roaming and non-roaming neutral fragmentation channel, we have plotted in Fig.~\ref{fig:3BodyIsoPropanol} the relevant dynamics from \ce{CH3+} + \ce{H2O} neutral + \ce{C2H3+} in 2-propanol (\ce{CH3CHOHCH3}). The PIPICO channel, shown in Fig.~\ref{fig:3BodyIsoPropanol} (a), is for \ce{CH3+} measured in coincidence with \ce{C2H3+}, with a missing mass of \ce{H2O}. This channel involves two primary fragments, \ce{CH3+} and \ce{C2H5O+}, the latter of which subsequently dissociates into \ce{C2H3+} and neutral \ce{H2O}, with a slope that can be calculated as $-m_{C_2H_3^+}$/($m_{C_2H_3^+}$ + $m_{CH_3^+}$) = -0.643 ~\cite{Eland1991}. A black dashed line of slope -0.643 is overlaid on this PIPICO channel to demonstrate the agreement between the expected and experimentally-obtained slope. Fig.~\ref{fig:3BodyIsoPropanol} (b) shows the angular correlations between the momentum vectors of neutral \ce{H2O} and \ce{C2H3+} (blue line) and neutral \ce{H2O} and \ce{CH3+} (orange line). Both these distributions are asymmetrically shifted either towards or away from 0$^{\circ}$ implying stronger momentum correlation, further corroborating that the neutral fragmentation of \ce{H2O} proceeds through a secondary decay from \ce{C2H5O+}. This behavior is in sharp contrast to that of corresponding angular distributions observed for the roaming neutral \ce{D2} (shown to the right for comparison).  Additionally, Fig.~\ref{fig:3BodyIsoPropanol} (c) shows the Newton diagram for this channel over the first 200 fs pump-probe delay window. Here, the momentum vector of \ce{C2H3+} is fixed along the x-axis, while those of \ce{CH3+} and the reconstructed \ce{H2O} are plotted on the top and bottom halves, respectively. Although the ionic fragments, \ce{C2H3+} and \ce{CH3+}, share similar characteristics to their counterparts in deuterated acetonitrile (shown to the right), the neutral fragments are significantly different. Specifically, the reconstructed momentum of neutral \ce{H2O} is strongly directed towards \ce{C2H3+}, which fully supports that \ce{H2O} is produced through a secondary decay from \ce{C2H5O+}. On the other hand, the reconstructed momentum of neutral \ce{D2} displays no such preferential angularity and has a broad distribution centered at low momentum. 

\begin{figure}[H]
\centering
\includegraphics[scale=0.65]{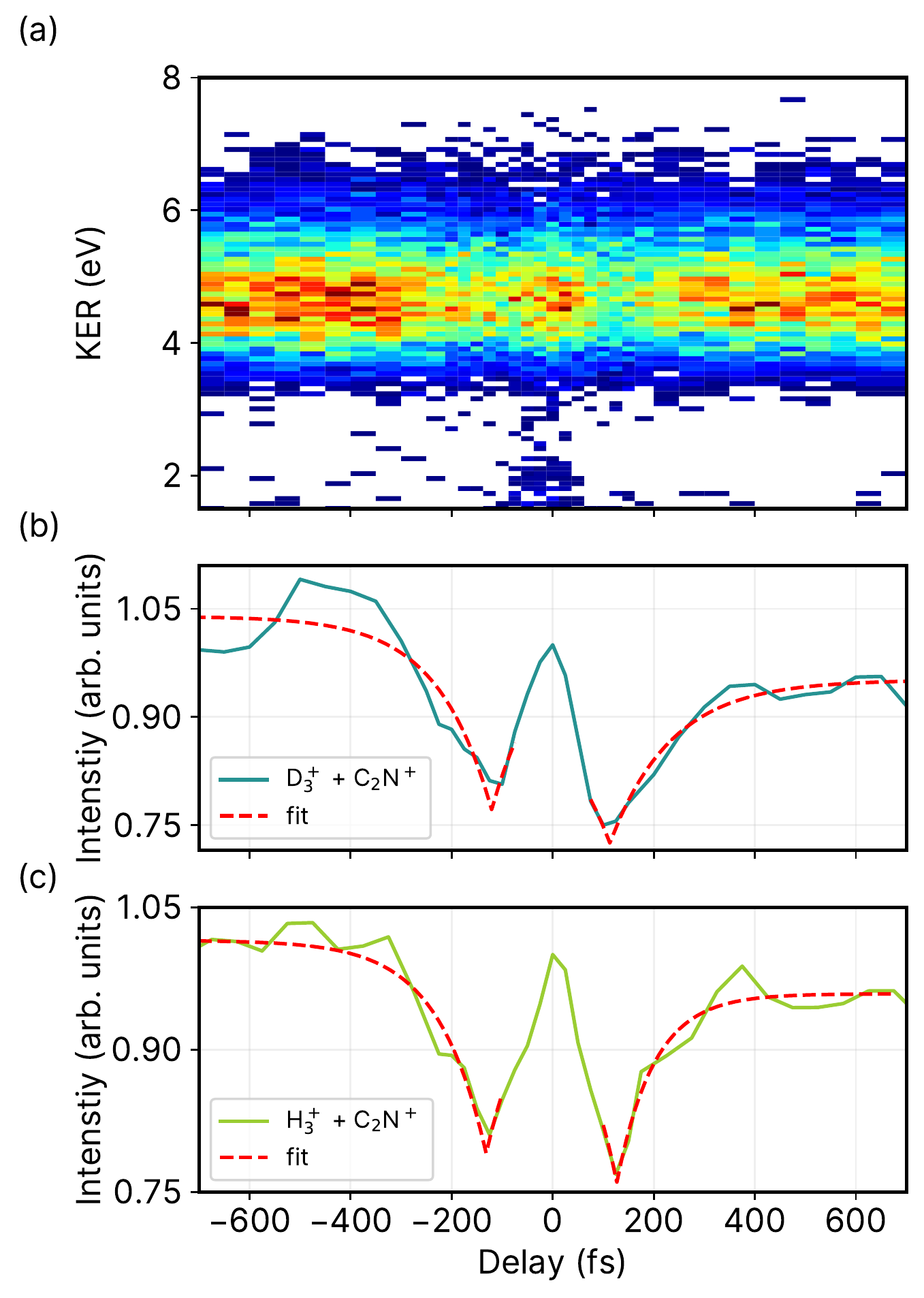}
\caption{(a) Kinetic energy release of \ce{D3+} + \ce{C2N+}. Projection of KER for (b) \ce{D3+} + \ce{C2N+} and (c) \ce{H3+} + \ce{C2N+} on the pump-probe delay axis. Red dashed lines obtained from the fit function (mentioned in the main text) are overlaid for comparison. The values from the fitting function are given in Table S1.}
\label{fig:DACNKER}
\end{figure}

\begin{table}[H]
    \begin{center}
    \caption{Fit values}
    \label{tab:fit_values}
        \begin{tabular}{c|c|c}
        
        \textbf{Signal} & \textbf{t$_0$} & \textbf{$\tau$}\\
        \hline
        \ce{H3+} Neg Delay & $-131\pm6$ fs & $111\pm23$ fs\\
        \ce{H3+} Pos Delay & $127\pm6$ fs& $75\pm20$ fs\\
        \ce{D3+} Neg Delay & $-121\pm10$ fs & $105\pm30$ fs\\
        \ce{D3+} Pos Delay & $113\pm5$ fs & $122\pm25$ fs\\
        
        \end{tabular}
    \end{center}
\end{table}

\begin{figure}[H]
\centering
\includegraphics[scale=0.375]{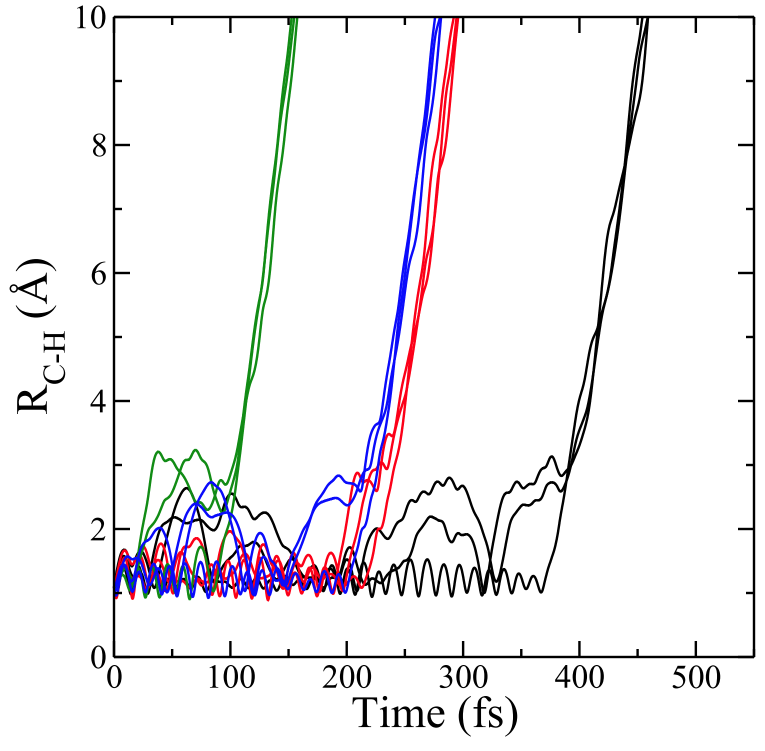}
\caption{Distance between the terminal carbon atom and each hydrogen atom (in \AA) as a function of the time in the trajectories of the molecular dynamics simulations leading to \ce{H_3^+}. The three C-H distances in each trajectory are given in different colors.}
\label{fig:CHDistances}
\end{figure}

\begin{figure}[H]
\centering
\includegraphics[width=0.45\linewidth]{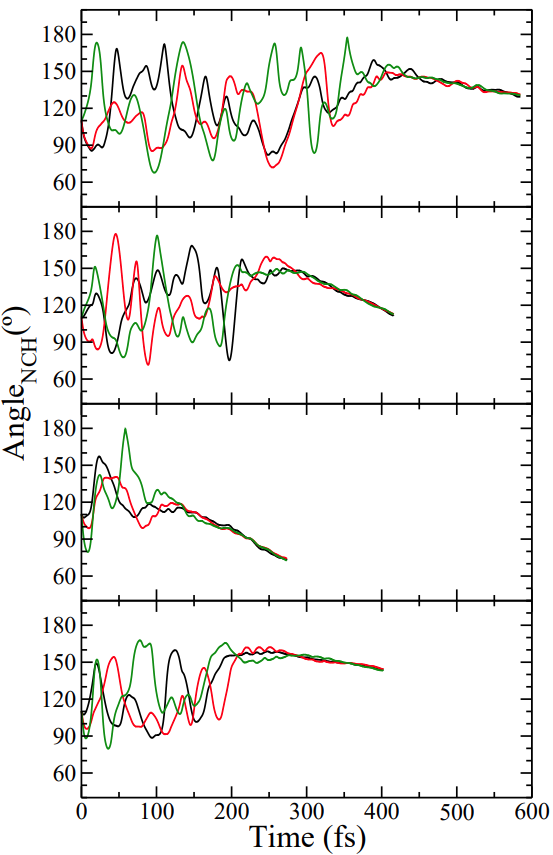}
\caption{Angle formed by the nitrogen, terminal carbon and hydrogen atoms (in degrees) as a function of the time in the trajectories of the molecular dynamics simulations leading to \ce{H_3^+}. The angle for each hydrogen atom is given in a different color in the trajectory.}
\label{fig:NCHANgle}
\end{figure}

\begin{figure}[H]
\centering
\includegraphics[scale=0.75]{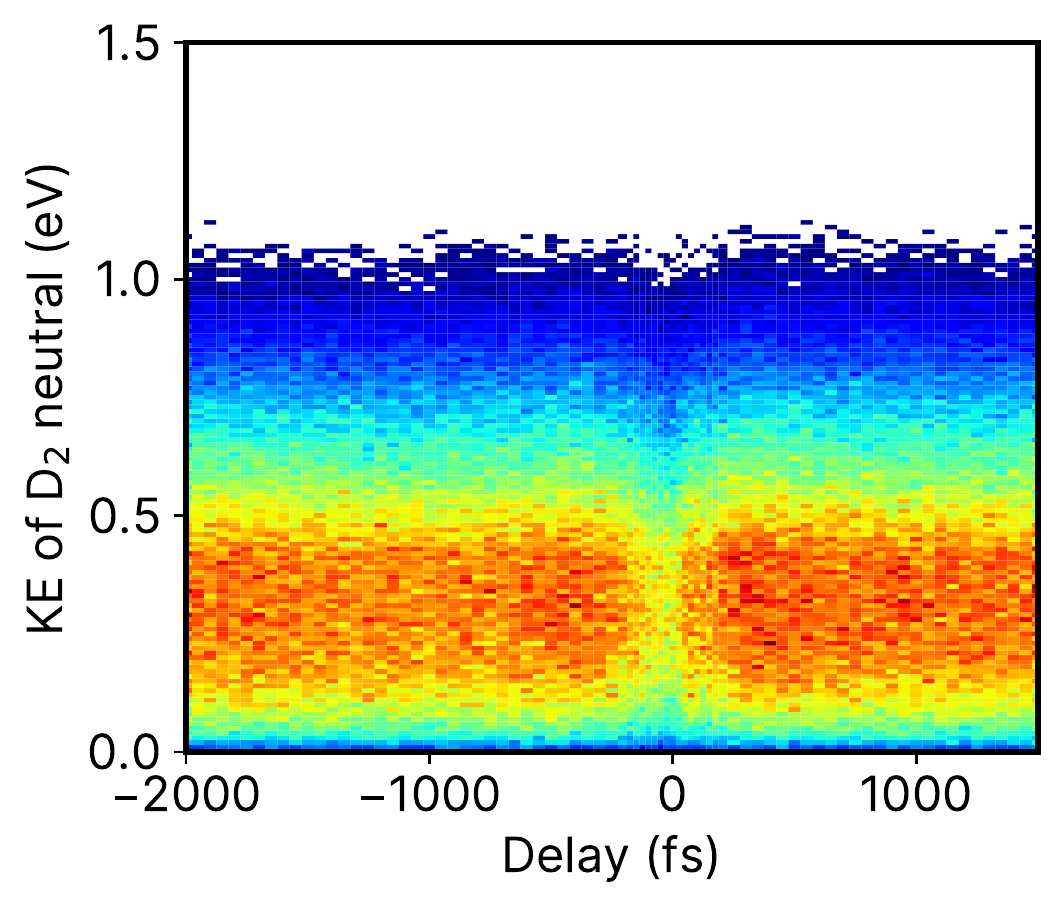}
\caption{Kinetic energy of the neutral \ce{D2} from the channel \ce{D+} + \ce{C2N+} + \ce{D2}.}
\label{fig:KE_D2Neutral}
\end{figure}

\begin{figure}[H]
\centering
\includegraphics[scale=0.75]{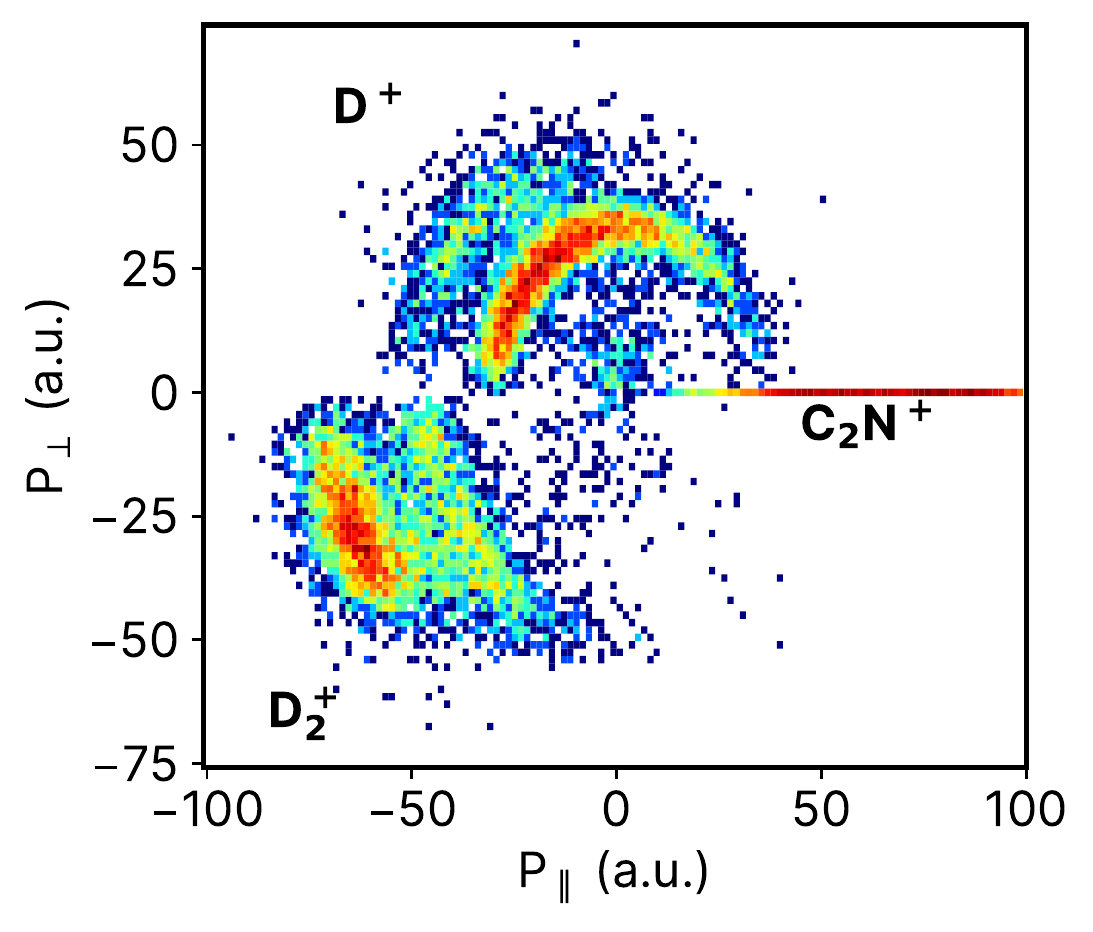}
\caption{Newton plots for the channel \ce{D+} + \ce{D2+} + \ce{C2N+} integrated over 200 fs time-delay window. The momentum vector of \ce{C2N+} is fixed along the x-axis and the momentum vectors of \ce{D+} and \ce{D2+} are plotted in the upper and lower halves of the plot, respectively.}
\label{fig:3BodyDACN_Newton}
\end{figure}

\begin{figure}[H]
\centering
\includegraphics[width=0.45\linewidth]{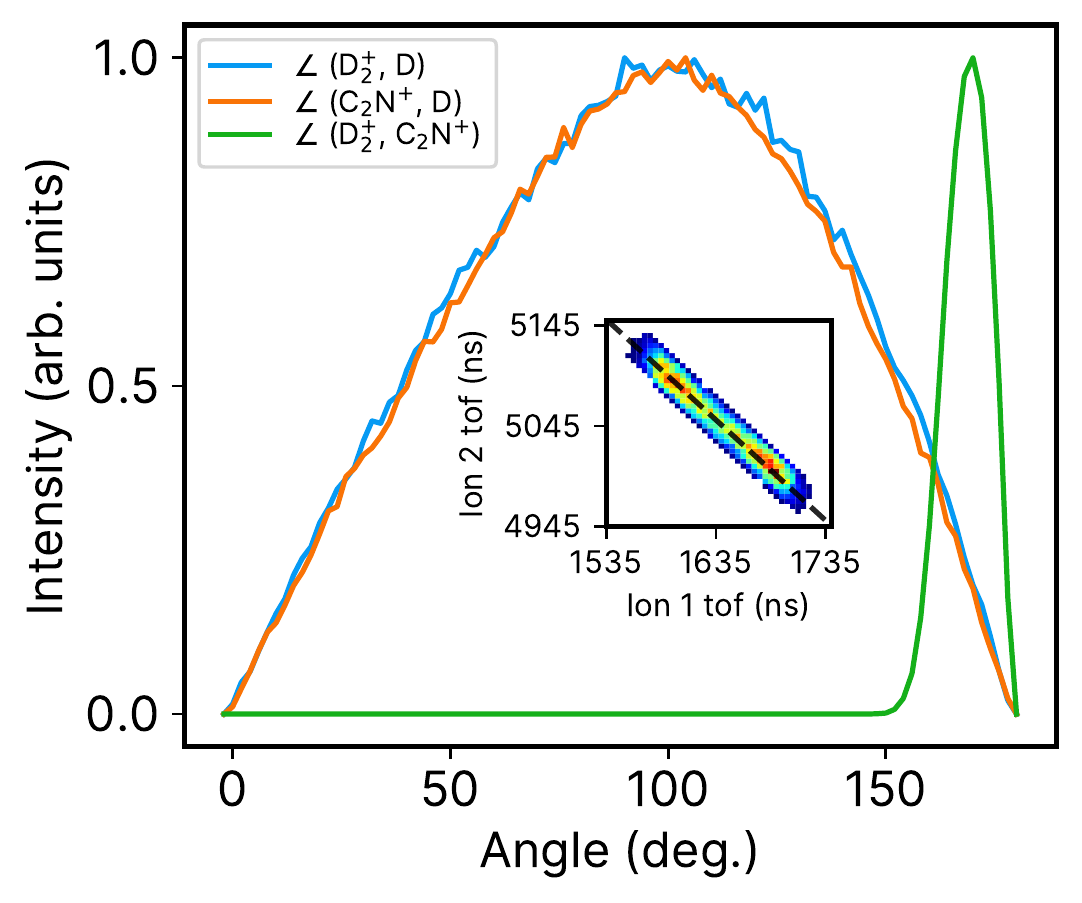}
\caption{Angular distributions between pairs of fragments in the incomplete \ce{D2+} + \ce{C2N+} + \ce{D} channel}
\label{fig:DRoaming}
\end{figure}

\begin{figure}[H]
\centering
\includegraphics[width=0.95\linewidth]{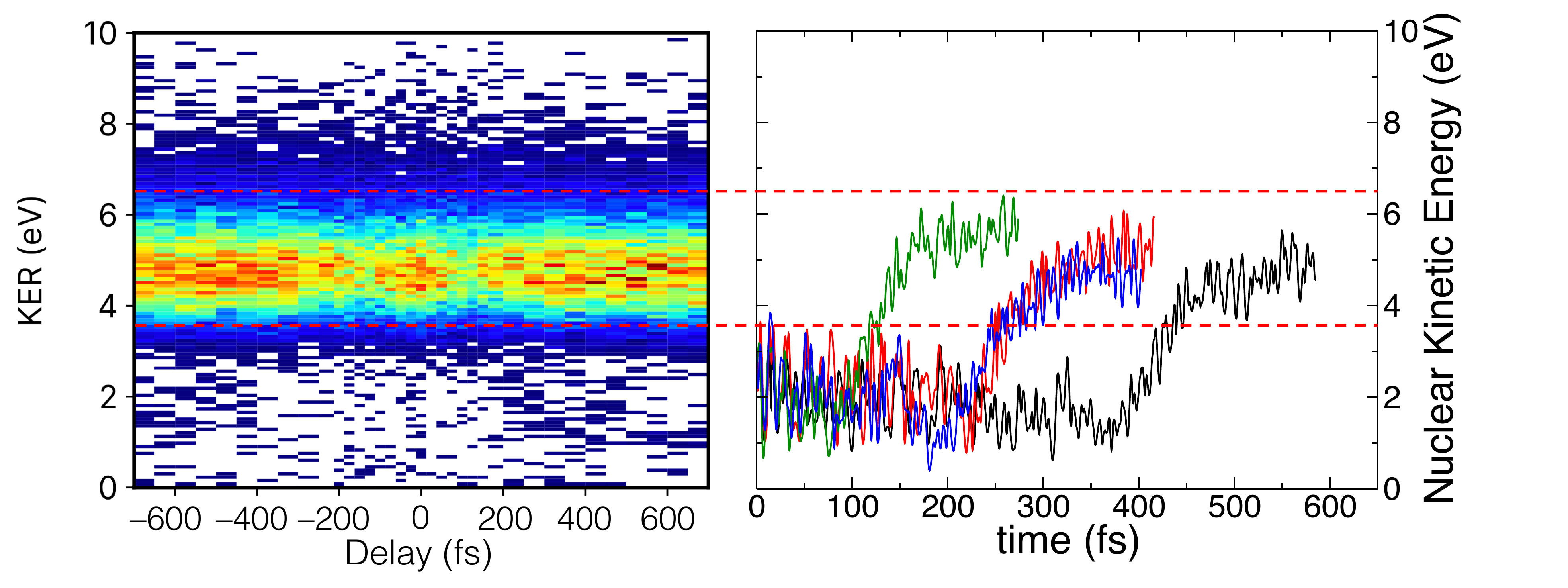}
\caption{Nuclear kinetic energy as a function of time for the four trajectories leading to \ce{H3+} + \ce{C2N+}.}
\label{fig:NKE_Simulations}
\end{figure}

\begin{figure}[H]
\centering
\includegraphics[width=0.85\linewidth]{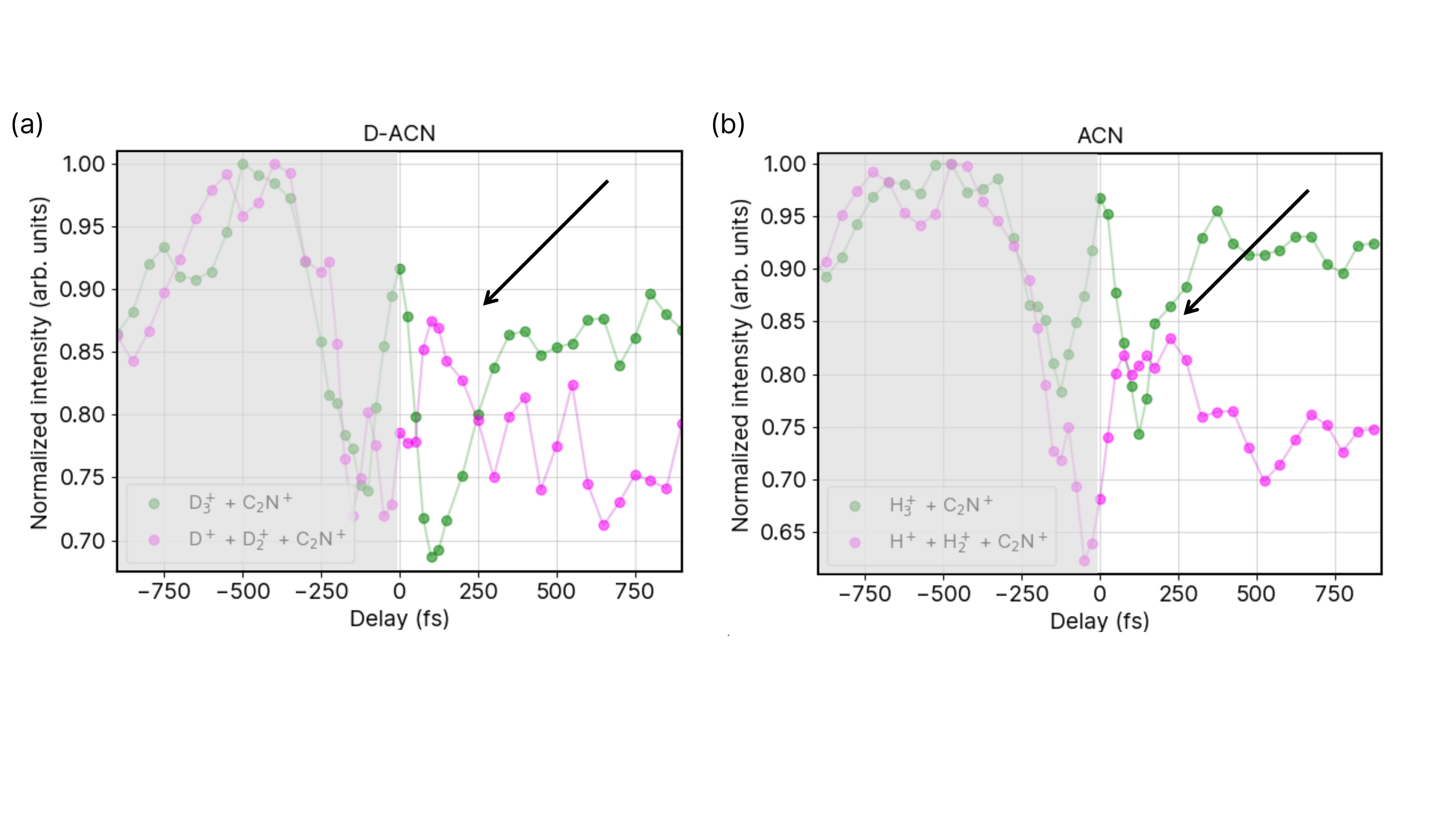}
\caption{Correlation between two-body (\ce{D3+} + \ce{C2N+} and three-body (\ce{D+} + \ce{D2+} + \ce{C2N+}) time-dependent KER dynamics in (a) deuterated acetonitrile and (b) acetonitrile. It is noted that the pump and probe pulses have slightly different beam diameters and pulse energies, and thus have different pulse intensities. For this measurement, positive delays have a stronger probe pulse than the pump pulse, and vice versa for negative delays. For positive delays,there is a correlated relationship between the double and triple ionization yields, which verifies that the depleted doubly ionized channel results in an enhancement of the triply ionized channel. For negative delays, both the double and triply ionized channels have the same time dependence. We believe this is due to the weaker probe pulse, which does not sufficiently ionize the neutral fragments.}
\label{fig:ThreeToTwoBody_Corr}
\end{figure}

\let\thefootnote\relax\footnotetext{* contributed equally}
\let\thefootnote\relax\footnotetext{$\dagger$ To whom correspondence should be addressed. Email: debadarshini.mishra@uconn.edu or aaron.laforge@uconn.edu}

\bibliography{acetonitrile}